\documentstyle{mn}

\newif\ifAMStwofonts
\def\mincir{\raise -2.truept\hbox{\rlap{\hbox{$\sim$}}\raise5.truept \hbox{$<$}\ }}
\def\mincireq{\hbox{\raise0.5ex\hbox{$<\lower1.06ex\hbox{$\kern-1.07em{\sim}$}$}}}
\def\magcir{\raise-2.truept\hbox{\rlap{\hbox{$\sim$}}\raise5.truept \hbox{$>$}\ }}

\title{Probing star formation with galactic cosmic rays}

\title{Probing star formation with galactic cosmic rays}

\author[Persic \& Rephaeli]
       {Massimo Persic$^{1}$, 
        Yoel Rephaeli$^{2,3}$\\
        $^1$INAF and INFN, via G.B.Tiepolo 11, I-34143 Trieste, Italy \\
        $^2$School of Physics \& Astronomy, Tel Aviv University, Tel Aviv 69978, Israel \\
        $^3$Center for Astrophysics and Space Sciences, University of California at 
		San Diego, La Jolla, CA 92093, USA} 
\date{Accepted ... ... ... ... .
      Received ... ... ... ... ;
      in original form ... ... ... ...}

\pubyear{2008}

\begin{document}

\maketitle

\label{firstpage}

\begin{abstract}
Cosmic-ray energy densities in central regions of starburst galaxies,
as inferred from radio and $\gamma$-ray measurements of, respectively, 
non-thermal synchrotron and $\pi^o$-decay emission, are 
typically $U_{\rm p}={\cal O}(100)$\,eV\,cm$^{-3}$, i.e. 
typically at least an order of magnitude larger than near the Galactic center 
and in other non-very-actively star-forming galaxies. We first show that these very 
different energy-density levels reflect a similar disparity in the respective 
supernova (SN) rates in the two environments, which is not unexpected given the 
SN origin of (Galactic) energetic particles. As a consequence of this correspondence, 
we then demonstrate that there is partial quantitative evidence that the stellar 
initial mass function (IMF) in starburst nuclei has a low-mass truncation at 
$\sim$$2\,M_\odot$, as predicted by theoretical models of turbulent media, in 
contrast with the much smaller value of $0.1M_\odot$ that characterizes the 
low-mass cutoff of the stellar IMF in `normal' galactic environments.

\end{abstract} 

\begin{keywords}
Galaxies: cosmic rays -- Galaxies: gamma-ray -- Galaxies: spiral -- Galaxies: star formation
\end{keywords}

\maketitle
\markboth{Persic \& Rephaeli: CRs and star formation}{}

\section{Introduction}

In the nuclear regions of starburst galaxies active star formation powers 
emission of radiation directly by supernova (SN) explosions and indirectly by 
SN-shock heating of interstellar gas and dust, as well as from radiative 
processes involving SN-shock--accelerated electrons and protons. 

Timescales of starburst activity in galaxies are comparable to 
galactic dynamical timescales, $\tau_{\rm SB} \sim \tau_{\rm dyn} 
\sim 10^8$\,yr. On the other hand, the timescales required for cosmic ray 
(CR) protons to gain energy by the Fermi-I acceleration process and to lose 
it (via pion decay into photons and $e^+e^-$ pairs, or advection) are 
$\tau_+ \sim \tau_- \sim 10^5$\, yr, respectively. 
Indeed, assuming the Fermi-I process to be at work in a SN remnant, $\tau_+ 
\equiv E/ \dot{E} = (\Delta E / E)^{-1} \Delta t = \beta^{-1} \Delta t = 
(10/\beta_{0.1})\, \Delta t \sim 10^5$\,yr, where $\Delta t \sim 10^4$\,yr 
is a typical SN remnant lifetime, and $\beta_{0.1}$ is the speed of the SN 
ejecta in units of $0.1\,c$. The main energy-loss timescale is largely 
determined by the starburst-driven outflow from the starburst region, i.e., typically 
$\tau_- \sim \tau_{\rm out} \sim 10^5$\,yr (see Sect.\,4). Thus, $\tau_{+} 
\sim \tau_{-} << \tau_{\rm SB}$.

The relative lengths of these timescales suggest that in a starburst region a 
balance can roughly be achieved between energy gains and losses for galactic 
CRs during a typical burst of star formation. Under basic hydrostatic and 
virial equilibrium conditions in a galaxy, a minimum-energy configuration 
of the field and the CRs may be attained [but see Beck \& Krause (2005) for 
a critical view]. This implies that energy densities of particles and 
magnetic fields are roughly in equipartition (e.g., Longair 1994).

The equipartition assumption enables deduction of the CR proton energy 
density, $U_{\rm p}$ (essentially all the particle energy content), from 
the measured synchrotron radio emission (which can be observed relatively 
easily), and a theoretically expected ratio of the CR proton to electron 
energy density, thereby in principle circumventing the need for a second observable, 
such as the predicted  -- and only very recently detected 
(from the nearby starburst galaxies NGC\,253 and M\,82: Abdo et al. 
2009, Acciari et al. 2009, Acero et al. 2009) --
very-high-energy (VHE: $\geq$100\,GeV) 
$\gamma$-ray emission. This emission is largely from CR proton 
interactions with ambient gas protons, which produce neutral ($\pi^0$) 
and charged pions; $\pi^0$ decays into $\gamma$-rays. On the other 
hand, because in a starburst region $\tau_{-} << \tau_{\rm SB}$, $U_{\rm p}$ 
can be roughly estimated also from SN rates and the fraction of SN energy 
that is channeled into particle acceleration. 

In this paper we argue the the basic underlying connection between radiative 
yields of CR electrons and protons and SN rates, can be exploited to obtain some 
insight on the nature of the stellar initial mass function in the starburst region. In 
Assuming particle-field equipartition we derive (Sect.\,2) an approximate 
expression for $U_{\rm p}$ as a function of the radio flux.  
Using radio data from the literature, in Sect.\,3 we show that $U_{\rm p} 
\sim {\cal O}(100)$\,eV\,cm$^{-3}$ in the central starburst nuclei of three well-known 
actively star forming galaxies. For the same starburst galaxies, in Sect.\,4 we estimate 
$U_{\rm p}$ from the statistics of SN events and the (putative) SN energy 
that goes into particle acceleration. We show that these respective 
estimates are in reasonable agreement, even if appreciably uncertian. 
In Sect.\,5 we outline how this consistency can be used as a diagnostic of 
the initial stellar mass function (IMF), particularly of its low mass end.
Our conclusions are summarized in Sect.\,6.

\section{Particle and magnetic field energy densities}

Energetic electron populations consist of primary (directly accelerated) 
and secondary (produced via charged pion decays) electrons. Their combined  
spectral density distribution is assumed to be a single power-law of the 
Lorentz factor, $\gamma$, in some interval $\gamma_1 \leq \gamma \leq \gamma_2$ 
\begin{eqnarray}
N_{\rm e}(\gamma) ~=~ N_{e,{\rm 0}}\, (1+\chi) ~ \gamma^{-q} , & & 
\label{eq:el_spectrum}
\end{eqnarray}
where $N_{e,{\rm 0}}$ is a normalization factor of the primary electrons, and 
$\chi$ the secondary-to-primary electron ratio.

Electron synchrotron emission from a region with radius $r_{\rm s}$ (of an 
equivalent sphere that has the same volume) and mean magnetic field $B$, 
located at a distance $d$, is 
\begin{eqnarray}
\lefteqn{
f_\nu ~=~  ~ 5.67 \times 10^{-22}~ { r_{\rm s}^3 \over d^2}  
N_{\rm e,0} (1+\chi) ~a(q) ~B^{q+1 \over 2} ~ \times ~}
                \nonumber\\
& & {} ~~~~~~~~~~~~~ \times ~ \bigg({\nu \over 4 \times 10^6}\biggr)^{-{q-1 \over 2}} 
~~~ {\rm erg/(s~ cm^2 Hz)} ,
\label{eq:synchr}
\end{eqnarray}
where $a(q)$ is defined and tabulated in, e.g., Tucker (1975). 
Setting $\psi \equiv ({r_{\rm s}/0.1\,{\rm kpc}})^{-3}\,({d /{\rm Mpc}})^
2\,( {f_{1\,{\rm GHz}}/{\rm Jy}})$ and $\nu=1$\,GHz, 
the the normalization of the electron spectrum (Eq.(\ref{eq:synchr}) is 
\begin{eqnarray}
N_{\rm e,0} ~=~ {5.72 \over 1+\chi} \times 10^{-15} \, \psi ~ a(q)^{-1} 
B^{-{q+1 \over 2}} 250^{q-1 \over 2}
\label{eq:el_spect_norm}
\end{eqnarray}

A second relation is required to separately estimate $N_{\rm e}$ and $B$. 
This is provided by equipartition between particles (electrons and protons) 
and the magnetic field, $U_{\rm p} + U_{\rm e} \simeq B^2 / 8 \pi$, 
which could possibly be attained due to a high degree of coupling between all 
the relevant degrees of freedom in the starburst region. 
In terms of the proton to electron energy density ratio, $\kappa(q)/(1+\chi) = 
U_{\rm p}/U_{\rm e}$, equipartition translates to 
\begin{eqnarray}
U_{\rm p} \bigl[1+ {1+\chi \over \kappa(q)}\bigr] ~\simeq~ {B^2 \over 8\, \pi}\,. 
\label{eq:equip2}
\end{eqnarray}
The electron energy density is $U_{\rm e}$$\,=\,$$N_{\rm e,0}\,$$(1+\chi)\,$$m_{\rm e}c^2\,$ $
\int_{\gamma_1}^{\gamma_2} \gamma^{1-q} \,{\rm d}\gamma$.
An approximate expression for $U_{\rm e}$ is obtained by ignoring 
the contribution of low energy electrons with $\gamma < \gamma_1$, whose energy 
losses are dominated by Coulomb interactions with gas particles. While the change 
of the electron spectrum at low energies (when compared with that at higher 
energies, which is directly deduced from radio measurements) is important (e.g., 
Rephaeli 1979, Sarazin 1999), for our purposes here we can simplify by considering 
only the spectrum above $\gamma_1$. In the starburst regions of interest here $\gamma_1 
\sim 10^3$, but our numerical estimates depend only weakly on its exact value. 
Under this assumption, the electron energy density is $U_{\rm e} = N_{\rm e,0}\,(1+
\chi)\,m_{\rm e}c^2\,\int_{\gamma_1}^{\gamma_2} \gamma^{1-q} {\rm d}\gamma$, where 
$\gamma_2$ is an upper cutoff whose exact value is irrelevant in the applicable 
limit $\gamma_2 >> \gamma_1$. For $q>2$ and $\gamma_2 >> \gamma_1$, $U_{\rm e} \simeq 
N_{e,{\rm 0}}(1+\chi) m_{\rm e}c^2 \gamma_1^{2-q}/(q-2)$, which upon substitution of 
the expression for $N_{{\rm e},0} (1+\chi)$ from Eq.(\ref{eq:el_spect_norm}) yields 
\begin{eqnarray}
U_{\rm e} ~=~ {2.96 \over (1+\chi)} \times 10^{-22} \, 250^{q \over 2}\, \psi \, {\gamma_1^{-q+2} 
\over (q-2) ~a(q)}\, B^{-{q+1 \over 2}}\,.
\label{eq:el_en_dens}
\end{eqnarray}
Using Eq.(\ref{eq:equip2}), we then get the equipartition magnetic field 
\begin{eqnarray}
\lefteqn{ B_{\rm eq} ~=~ \biggl[ {7.44 \times 10^{-21} \over 1+\chi} \, \bigl[1+{\kappa(q) \over 
1+\chi}\bigr]\, {\gamma_1^{2-q}\, 250^{q/2} \, \psi \over (q-2)\, a(q)} \biggr]^{2 \over 5+q}  
\,. }
\label{eq:equip_B}
\end{eqnarray}
Finally, using Eqs.(\ref{eq:equip_B}),(\ref{eq:equip2}) and the expression for $\kappa(q)$ (e.g., 
Persic et al. 2008), we readily obtain an explicit expression for $U_{\rm p}$. 

%%%%%%%%%%%%%%%%%%%%%%%%%%%%%%%% TAB 1 %%%%%%%%%%%%%%%%%%%%%%%%%%%%%%%%%%%%%%%
\begin{table}
\caption[] {Star-forming galaxies: {\it IRAS} data.}
%\begin{flushleft}
\begin{tabular}{ l  l  l  l  l  l  l }
%\noalign{\smallskip}
\hline
\hline
\noalign{\smallskip}
Object & d$^{(a)}$ & $f_{12\mu}^{(b)}$ & $f_{25\mu}^{(b)}$ & $f_{60\mu}^{(b)}$ & $f_{100\mu}^{(b)}$  & 
	$L_{\rm TIR}^{(c)}$\\
\noalign{\smallskip}
\hline
\noalign{\smallskip}
NGC\,253              & 3.0  & 41.04& 154.67&  967.81& 1288.15 & 44.00\\
M\,82                 & 3.6  & 79.43& 332.63& 1480.42& 1373.69 & 44.35\\
Arp\,220              & 72.3 &  0.61&   8.00&  104.0 & ~112.0  & 45.68\\
\noalign{\smallskip}

\hline
\end{tabular}
%\end{flushleft}
\smallskip

$^{(a)}$ Distances in Mpc. 
$^{(b)}$ {\it IRAS} flux densities (in Jy), from Genzel et al. (1998). 
$^{(c)}$ TIR luminosities (in erg s$^{-1}$), in log form. 

\end{table}
%%%%%%%%%%%%%%%%%%%%%%%%%%%%%%%%%%%%%%%%%%%%%%%%%%%%%%%%%%%%%%%%%%%%%%%%%%%%%%

\subsection{Secondary electrons}

The ratio $\chi$ is clearly a function of energy; it depends on the injection 
proton-to-electron number ratio, $p/e$, which generally depends on the injection 
slope (e.g., Schlickeiser 2002), and on the gas optical thickness to {\it pp} 
interactions (that produce neutral and charged pions). Because of the high gas 
density in the starburst region the production of secondary electrons is expected to be 
relatively important.
 
Each $e^+$ resulting from $\pi^+$ decay (following a pp interaction) generates 
(through a photon pair obtained by interaction with an ambient $e^-$) two $e^+e^-$ 
pairs. The $e^-$'s from the pairs are secondary electrons whose energy is 
${1 \over 4}$ that of the initial $e^+$. The positrons in the pairs repeat the 
loop. This cascade continues as long as the photon energy is not smaller than 
$1.022$\,MeV. However, for magnetic fields $\sim$$100\,\mu$G (see below) and 
an electron density of $n_{\rm e}$, this low threshold value is replaced by a 
higher threshold, 
$E_{\rm c}$, below which electrons lose energy much more rapidly by Coulomb 
interactions with gas particles than by synchrotron emission (e.g., Rephaeli 
1979). With typical gas densities $\sim$100\,cm$^{-3}$ in the central region 
of a starburst galaxy (such as M\,82 and NGC\,253, which we specifically consider 
in the next section), $E_{\rm c}$$\approx$1\,GeV. In our approximate treatment 
here we use $E_{\rm c}$ as the threshold energy for the production of secondary 
electrons. 

In order to evaluate the number of secondary electrons produced by 
a power-law distribution of CR protons where every proton undergoes 
one {\it pp} interaction, let us consider that a proton of kinetic 
energy $T$ produces secondary electrons with energies successively 
lower by a factor 4, starting at ${T / 3}$ and down to $E_{\rm c}$. 

In our simple shower model the number of secondary electrons produced 
per interacting proton of kinetic energy $T$ -- i.e. those produced down 
to the last step of the shower, k$_{\rm max}$$=$$[{\rm log}(3\,E_{\rm 
c}T^{-1})/{\rm log}(1/4)]\,+\,1$ -- is $2^{k_{\rm max} -1}$. (The first 
step of the shower, k$=$1, corresponds to the first $e^+$ resulting from 
$\pi^+$ decay following the initial pp interaction.) Integrating over 
a $\propto T^{-q}$ CR proton spectrum and multiplying by the relevant  
injection p/e ratio, we obtain the secondary-to-primary electron ratio 
corresponding to the case when all CR protons produce secondary electrons, 
$\chi_0$$=$$r_{p/e}\int_{E_{\rm c}}^{E_{\rm in}} 2^{[k_{\rm max}(T)-1]} 
T^{-q}\, {\rm d}T$, where $r_{p/e}$ is the proton to electron density ratio.
For a strong non-relativistic shock ($q$$=$2, implying $p/e$$=$43), $E_{\rm 
c}$$=$1\,GeV, and $E_{\rm in}$$=$10\,TeV we get $\chi_0$$\simeq$50. 
However, not every CR proton produces secondary electrons. The mean free 
path of CR protons in a medium of density $n_{\rm p}$ due to {\it pp} 
interactions is $\lambda_{\rm pp} = (\sigma_{\rm pp} n_{\rm p})^{-1}$; 
for protons with kinetic energy $T\,\sim$\,{\rm few}\,TeV the cross section 
is $\sigma_{\rm pp} \simeq 50$\,mb. For an ambient gas density $n_{\rm p} 
\simeq 100$\,cm$^{-3}$, $\lambda_{\rm pp} \sim 65$\,kpc. The probability 
for a single CR proton to undergo a {\rm pp} interaction in its 3D random 
walk through a region with a typical radial extent $r_{\rm s}$$\approx$1/4\,kpc 
is then $\sqrt{3} \,r_{\rm s} / \lambda_{\rm pp} \simeq 6.7 \times 10^{-3}$.
We then estimate that for a strong non-relativistic shock the secondary to 
primary electron ratio is $\chi$$=$$\chi_0\, \sqrt{3}\,(r_{\rm s}/\lambda_{\rm 
pp})$$\simeq$$0.33$. 

In the rest of this paper we shall use $\chi$$=$0.5 as a fiducial value. 
This is in approximate agreement with results of detailed numerical starburst 
models for energies $\magcir$10\,MeV (e.g., Rephaeli et al. 2009; see 
also Paglione et al. 1996; Torres 2004; Domingo-Santamar{\'\i}a \& Torres 
2005; De Cea et al. 2009).

\section{CR energy density in star-forming environments}

In the previous section we saw that if the source size, distance, and 
radio flux and spectral index are known, then the corresponding particle 
energy density can be evaluated. As an application, in this section we 
compute the particle energy densities in the starburst nuclei of three nearby 
star-forming galaxies with different current star formation rates (SFRs): 
the two nearby galaxies, NGC\,253 and M\,82, and the merging starburst Arp\,220 
(SFR$=5$, $10$, and $220\,M_\odot$\,yr$^{-1}$, respectively: see Table\,1 
and Eqs.(\ref{eq:kennicutt}),(\ref{eq:TIR})). These estimates are then 
compared with the corresponding mean value in the Galaxy [SFR\,$\simeq 
(1-2)\,M_\odot$\,yr$^{-1}$]. 

For NGC\,253 (distance $d=3$ Mpc), following Rephaeli et al. (2009), we 
assume the following parameters for the central starburst emission: $r_{\rm s}=
200$\,pc, $f_{\rm 1\,GHz}=5.6$\,Jy, and $\alpha \simeq 0.75$. Assuming 
that this radio flux is synchrotron emission from a non-thermal population 
of energetic electrons, the spectral index is $q$$\simeq$$2.5$, hence 
$a$$=$$0.0852$ and $\kappa$$\simeq$$8$. Setting $\chi$$=$$0.5$, we obtain 
$U_{\rm p}$$\approx$$75$\,eV\,cm$^{-3}$. 

For M\,82 we adopt (see Persic et al. 2008) $d=3.6$ Mpc, $r_{\rm s}=260$\,pc, 
$f_{\rm 1\,GHz}=10$\,Jy, $\alpha$$\simeq$$0.71$, and $q$$\simeq$$2.42$ (the 
latter implies $a \simeq 0.09$ and $\kappa \simeq 10$). Setting $\chi$$=$$0.5$, 
we obtain $U_{\rm p}$$\approx$97\,eV\,cm$^{-3}$.

In Arp\,220 (distance $d$$=$$72$ Mpc), the starburst activity takes place in a central 
molecular-gas disk of radius 480\,pc and thickness 90\,pc that accounts for 
SFR$\sim$$120\,M_\odot$\,yr$^{-1}$, and embeds (at a galactocentric distance 
of 200\,pc) two extreme starburst nuclei of radii 68 and 110 pc that account for 
SFR$=50$ and 35\,$M_\odot$\,yr$^{-1}$, respectively. This scheme [derived 
from Torres (2004) and Eq.(\ref{eq:kennicutt}) below] is consistent with the 
galaxy-wide TIR-inferred SFR\,$\sim 225\,M_\odot$\,yr$^{-1}$. For use in 
Eq.(\ref{eq:equip_B}), for the disk {\it alone} we assume an effective 
spherical radius $r_{\rm s}\simeq 250$\,pc, $f_{\rm 1\,GHz}\simeq 0.3$\,Jy, 
and $\alpha \simeq 0.65$: for $\chi=0.5$ [see Torres (2004), the bottom of 
his Fig.7, 
for $N_{\rm p}/N_{\rm e}=100$, consistent with $q=2.3$ (Schlickeiser 2002) 
-- the latter value implies $a \simeq 0.09$ and $\kappa \approx 12$], this 
leads to $U_{\rm p} \approx 520$\,eV\,cm$^{-3}$. The two extreme starburst nuclei 
embedded in the disk are environments that are probably similar to very young 
SN remnants, and hence far from equipartition (e.g., Torres 2004).

By contrast, in the Milky Way -- a much more sedate star-forming environment 
with SFR\,$\sim (1-2)$\,$M_\odot$ -- the value of the proton energy density is 
$U_{\rm p}$$\sim$$1$\,eV\,cm$^{-3}$ at the Earth's position (see Schlickeiser 
2002), and $(6 \pm 3)$\,eV\,cm$^{-3}$ in the $\sim$200\,pc region of the 
Galactic center, as inferred from the measured VHE $\gamma$-ray emission 
(Aharonian et al. 2006). In the even more sedate Large Magellanic Cloud 
(LMC), $U_{\rm p}$$\sim$$(0.2-0.3)$\,eV\,cm$^{-3}$ as inferred from the 
measured HE emission (Kn\"odlseder 2009).

\section{Cosmic rays and supernova rates}
\subsection{Estimates}

Combining the SN frequency with the residence timescale of energetic protons 
that give rise to TeV emission in the star-forming region, and assuming a 
bona-fide value of the energy that goes into accelerating CR particles per 
SN event, we can obtain a second, independent estimate of $U_{\rm p}$.  
Moroever, as we show below this estimate for $U_{\rm p}$ agrees well with 
that deduced for each of the four galaxies in Sect.\,3.  
This match does hint a link between cosmic rays and SNe, in agreement with 
previous suggestions (Ginzburg \& Syrovatskii 1964; Salvati \& Sacco 2008).

The energy-loss timescale for {\it pp} interactions is $\tau_{\rm pp} = 
(\sigma_{\rm pp} c n_{\rm p})^{-1}$. For protons with kinetic energy $T \sim 
{\it few}$\,$\times$\,10 TeV\, $\sigma_{\rm pp} \simeq 50$\,mb (hardly varying 
in the energy range of interest). Hence, 
\begin{eqnarray}
\lefteqn{ 
\tau_{\rm pp} ~\sim~ 2 \times 10^7 ~ n_{\rm p}^{-1}~~~~ {\rm yr} \,.
}
\label{eq:pp_time}
\end{eqnarray}
In the $r_{\rm s} = 0.3$\,kpc central starburst of M\,82 the gas mass is $M \simeq 2 
\times 10^8\,M_\odot$ (Persic et al. 2008, and references therein), so $n_{\rm p}  
\simeq 100$\,cm$^{-3}$; therefore,  
$\tau_{\rm pp} \sim 2 \times 10^5$\,yr for ${\cal O}(10)$ TeV proton CRs in the 
central starburst of M\,82. A similar estimate for 
NGC\,253 [with $r_{\rm s} = 0.2$\,kpc (Forbes \& DePoy 1992), $M_{H_2} = 
7.2 \times 10^7\,M_\odot$ (Mauersberger et al. 1996)] 
yields very similar value, $\tau_{\rm pp} \sim 10^5$\,yr. 
In the star-forming disk of Arp\,220,  $\tau_{\rm pp} \sim 1.7 \times 10^4$\,yr 
(based on the gas density quoted in Table 2 of Torres 2004). 

However, for the galaxies considered in this paper the $\pi$-production timescale 
is longer than the CR proton outflow timescale, $v_{\rm out}$. The latter is 
dominated by advection of the energetic particles out of the disk mid-plane 
region in a fast starburst-driven wind. For M\,82, $v_{\rm out} 
\sim 2500$\,km\,s$^{-1}$ (Chevalier \& Clegg 1985; Seaquist \& Odegard 
1991; Strickland \& Heckman 2009). Assuming a homogeneous distribution of 
SNe within the starburst nucleus of radius $r_{\rm s}$, the outflow timescale is then 
\begin{eqnarray}
\lefteqn{ 
\tau_{\rm out} 
~=~ 3 \times 10^4 ~ \bigl({r_{\rm s} \over 0.3\,{\rm 
kpc}}\bigr) ~ \bigl({ v_{\rm out} \over 2500\,{\rm km\,s}^{-1}}\bigr)^{-1}~ {\rm yr}\,. 
}
\label{eq:out_time}
\end{eqnarray}

During the timescale in which protons actually reside in the central starburst region, 
namely the shorter of the latter two timescales, $\tau_{\rm out}$, a number $\nu_{\rm 
SN}$$\tau_{\rm out}$ of SN explode and deposit the kinetic energy of their 
ejecta, $E_{\rm ej}$$=$$10^{51}$\,erg (Woosley \& Weaver 1995), into the 
interstellar medium. Arguments based on the CR energy budget in the Galaxy 
and SN statistics suggest that a fraction $\eta$$\sim$$0.05-0.1$ of 
this energy is available for accelerating particles (e.g., Higdon et al. 1998; 
Tatischeff 2008). We then express the CR proton energy density as: 
\begin{eqnarray}
\lefteqn{
U_{\rm p} ~ = ~ 85 ~ {\nu_{\rm SN} \over 0.3\,{\rm yr^{-1}}}  ~ 
{\tau_{\rm out} 
\over 3 \times 10^4\,{\rm yr}}   ~ {\eta \over 0.05}~ {E_{\rm ej} \over 
10^{51}\,{\rm erg}} ~ \bigl( {r_{\rm s} \over 0.3\,{\rm kpc}} \bigr)^{-3} } 
\nonumber\\
& & {} ~~~~~~~~~~~~~~~~~~~~~~~~~~~~~~~~~~~~~~~ {\rm eV~ cm}^{-3}.
\label{eq:CRp_density}
\end{eqnarray}

From Eqs.(\ref{eq:CRp_density}),(\ref{eq:SNrate3}) we see that a substantial 
agreement with our equipartition estimates in Sect.\,3 is reached for our five 
galaxies: 
\smallskip

\noindent 
{\sl NGC\,253:} 
Taking $r_{\rm s}$$=$$0.20$\,kpc (Forbes \& DePoy 1992) and $\nu_{\rm SN}$$=$$0.12$\,yr$^{-1}$ 
(Rieke et al. 1988)), we get $U_{\rm p}$$\simeq$$75$\,eV\,cm$^{-3}$;
\smallskip

\noindent
{\sl M\,82:} Taking 
$r_{\rm s}$$=$$0.26$\,kpc (Persic et al. 2008 and references therein) and $\nu_{\rm 
SN}$$=$$0.25$\,yr$^{-1}$ (Rieke et al. 1980), we get $U_{\rm p}$$\simeq$$95$\,eV\,cm$^{-3}$; 
\smallskip

\noindent
{\sl Arp\,220:} 
Taking $r_{\rm s}$$=$$0.25$\,kpc (see Sect.\,3) in Eq.(\ref{eq:CRp_density}) and 
$0.09$\,kpc (i.e., the thickness of the starburst disk, perpendicular to which the wind 
will break out) in Eq.(\ref{eq:out_time}) and $\nu_{\rm SN}$$\approx$$3.5$\,yr$^{-1}$ 
(Smith et al. 1998; Lonsdale et al. 2006), we get $U_{\rm p}$$\simeq$$505$\,eV\,cm$^{-3}$. 
\smallskip

\noindent
{\sl Milky Way:} for the relatively quiet star forming environment of the Galaxy 
(SFR\,$\sim\,(1-2)\,M_\odot$\,yr$^{-1}$) we consider a nuclear star-formation 
volume comparable to those considered for the previous galaxies, $r_{\rm s} = 
0.2$\,kpc. The large-scale distribution of the HI gas in the Galaxy is made up 
of two major components (Kalberla \& Dedes 2008): most prominent is an exponential 
HI disk with $R_d=3.75$\,kpc, coplanar to the stellar disk and that can be traced 
out to 35\,kpc. Surrounding this region there is a patchy distribution of highly 
turbulent gas that reaches higher scale-heights and larger radial distances. If 
$n_{\rm HI}= 1$\,cm$^{-3}$ in the solar neighborhood, i.e. at a Galactocentric 
distance $R_0=8$\,kpc (Dehnen \& Binney 1998), the average value within $r_{\rm 
s}$ is $\bar n_{\rm HI} \simeq 8$\,cm$^{-3}$. Hence the relevant timescale to be 
used in Eq.(\ref{eq:CRp_density}) is $\tau_{\rm pp} = 2.5 \times 10^6$\,yr. The 
full disk Galactic SN rate, $\nu_{\rm SN}= 0.02$\,yr$^{-1}$ (Diehl et al. 2006), 
should be rescaled to the $r_{\rm s}$ region: assuming an exponential stellar 
disk with $R_d=2.5$\,kpc, (Binney et al. 1997 from COBE $L$-band data; see also 
Dehnen \& Binney 1998), the mass contained within $r_{\rm s}$ is $3.66 \times 
10^{-3}$ of that contained within $R_0$, hence the resulting SN 
rate to be used in Eq.(\ref{eq:CRp_density}) is $0.73 \times 10^{-4}$\,yr$^{-1}$. 
Assuming $\eta=0.05$, from Eq.(\ref{eq:CRp_density}) we finally obtain $U_{\rm p} 
\simeq 5.7$\,eV\,cm$^{-3}$. This value is fully consistent with that derived, 
for the same region, by Aharonian et al. (2006) from the diffuse VHE\,$\gamma$-ray 
emission measured there with the H.E.S.S. Cherenkov array. 
\smallskip

\noindent
{\sl LMC:} This satellite of the Milky Way has a very 
quiet star forming environment, characterized by $\nu_{\rm SN} = 0.002$\,yr$^{-1}$ 
(van den Bergh \& Tammann 1991) in the star forming region of the galaxy. The latter 
can be modeled as a truncated disk/spheroid with $r_{\rm t}$$\approx$3\,kpc 
whose thickness is also $\approx$3\,kpc (Weinberg \& Nikolaev 2001), we use 
$r_{\rm s}=3$\,kpc in Eq.(\ref{eq:CRp_density}). There is no mass outflow 
from the LMC, hence the energy-loss timescale for {\it pp} interaction is 
relevant here; with an average gas density of $n_{\rm p}$$\approx$2\,cm$^{-3}$, 
this timescale 
is $\tau_-$$=$$\tau_{\rm pp}$$\approx$$10^7$yr. Using these values in 
Eq.(\ref{eq:CRp_density}), we obtain $U_{\rm p}$$\approx$0.2\,eV\,cm$^{-3}$. 
This value is consistent with that inferred from {\it Fermi}/LAT measurements 
of the average emissivity spectrum of the LMC (either including or excluding 
the bright star-forming region of 30\,Doradus), which suggest a $U_{\rm p}$ 
value about 0.2-0.3 times that in the solar neighbourhood (Kn\"odlseder 2009). 

In summary, the CR energy densities estimated in five galactic nuclei of similar 
size (three starburst galaxies, the central Galactic region, and the LMC), appear to be largely correlated 
with key features of the ongoing star formation: 
local ambient density, SN rate, and CR proton residence time. In fact, for same $\eta E_{\rm ej}$ (by 
assumption) 
and similar $r_{\rm s}$ (from observations), the CR energy density $U_{\rm p}$ seems 
to be well described just as a function of the number of SN explosions during the 
CR residence timescale, $\nu_{\rm SN} \tau_{\rm res}$ , where the CR residence 
(energy-loss) timescale $\tau_{\rm res}^{-1}$ is defined in terms of the geometric 
mean of the {\it pp} interaction timescale and of the advection timescale, 
$\tau_{\rm res}^{-1} = \tau_{\rm pp}^{-1}(n_{\rm HI}) + \tau_{\rm out}^{-1}
(r_{\rm s})$. 

\subsection{Uncertainties}

Uncertainties in the quantities appearing in Eq.(\ref{eq:CRp_density}) need to be 
assessed before we make deduction about the stellar IMF.
\smallskip

\noindent 
$\bullet$ The values of $U_{\rm p}$ derived in Section 3, based on {\it particle-field 
equipartition} and representing mean values for the respective central starburst regions, are 
essentially confirmed by accurate numerical treatments based on the solution of the 
diffusion-loss equation for the accelerated particles (M\,82: Persic et al. 2008, and De 
Cea et al. 2009; NGC\,253: Paglione et al. 1996, Domingo-Santamar{\'\i}a \& Torres 2005, 
and Rephaeli et al. 2009; Arp\,220: Torres 2004). Starting with the initial accelerated 
spectrum in the central starburst region, these authors followed the evolution of the spectrum 
as the particles diffuse and are convected from this inner region to the outer disk (and 
halo). Including all the relevant leptonic and hadronic interactions in a numerical 
treatment, the particle spatial and energy distributions, and their respective radiative 
yields, were followed across the galaxy 
\footnote{Only in the two extreme starburst nuclei of Arp\,220 -- which we 
	don't consider in this paper -- is particles/field equipartition 
	most likely not 
attained. These strongly magnetized regions (where typically the mean field is at the mG 
level) are similar to very young SN remnants, and hence perhaps still too 'hot' for any 
sort of equipartition to be attained (Torres 2004). }.
  
At the time of this writing $\gamma$-ray detections of M\,82 and NGC\,253 in the HE band 
(Abdo et al. 2009) and in the VHE band (Acciari et al. 2009; Acero et al. 2009) were 
announced. The measured fluxes and spectra of both galaxies in the two bands agree with 
the predictions of recent numerical models (Persic et al. 2008; de Cea Del Pozo et al. 2009; 
Paglione et al. 1996; Domingo-Santamar{\'\i}a \& Torres 2005; Rephaeli et al. 2009), which 
are all based on values of $U_{\rm p}$ that are in agreement, to within a factor of 
$\approx$2, with our estimates in Sect.4.  

Given these arguments, we feel confident that the energy densities of CR protons in starburst 
nuclei, as derived from our equipartition arguments above, are reliable to within a factor 
of $\mincir$2.
\smallskip

\noindent 
$\bullet$ {\it SN rates in starburst nuclei} are in principle observable quantities. However, heavy 
optical extinction drastically reduces one's ability to track SN activity, whereas radio counts 
of SN remnants need information on their ages in order to be turned into actual SN rates. In 
general, long monitoring timescales, with measurements of actual source fadings and new 
appearances, are required to estimate the actual SN rate from the observation of the compact 
radio sources (Ulvestad \& Antonucci 1997). As for our three starburst galaxies, we assess (based on 
published observational results) that their SN rates are known to within a factor of $\mincir$1.5. 
\smallskip

\noindent 
$\bullet$ The uncertainty in the {\it CR proton outflow timescale}, $\tau_{\rm out}$, 
arises from the uncertainty in the fast wind velocity, $v_{\rm out}$. The latter is 
probably known to within $\sim$50\%. Indications of such velocity in M\,82 come from 
multifrquency radio measurements of its nonthermal halo emission (Seaquist \& Odegard 1991). 
The halo  emission is interpreted as synchrotron radiation from relativistic electrons 
accelerated in SN remnants and then swept out the disk plane by the extensive wind associated 
with the starburst (Chevalier \& Clegg 1985; Seaquist \& Odegard 1991). The steepening of the radio 
spectral index (from $-0.4 \pm 0.1$ in the starburst nucleus to about $-1$ in the halo), is explained 
by a simple model which involves outward convection of relativistic electrons which suffer 
energy losses by Compton scattering against IR photons (which are plentiful in the presence 
of the central starburst) and adiabatic expansion (which makes this loss term largely dominant in 
the outer radio halo). In the context of this model a convection speed $v_{\rm out} \sim 
(1000 - 3000)$\,km\,s$^{-1}$ is deduced, as well as a fraction of the total energy input to 
the halo in the form of relativistic particles of a few percent, consistent with the fraction 
found in Galactic SN remnants -- this agreement, too, supporting the hypothesis that the 
relativistic particles in the halo are produced by SN remnants in the disk (Seaquist \& 
Odegard 1991). The relationship between this fast particle outflow and the slower, denser 
minor-axis outflow observed at optical and X-ray wavelengths (with deprojected velocities 
of $\sim$$600$\,km\,s$^{-1}$) is unclear. Much of the optical data may be explained from 
large-scale shocks from the high-speed wind plowing into the gaseous halo and entrained disk 
gas (Shopbell \& Bland-Hawthorn 1998). 

\noindent
The assumed wind speeds are in agreement with the suggestion, from standard 
wind theory, that wind terminal velocities usually amount to some (small) 
multiple of the escape speed, $v_{\rm esc}$ (e.g., Parker 1958). For example, 
the terminal wind velocity may be $\approx$$(3-5)\,v_{\rm esc}$ in M\,82 
(Strickland \& Heckman 2009), and $\approx$$3\,v_{\rm esc}$ in our 
Galaxy (Everett et al. 2008). Would the correspondingly short convective 
timescales imply similarly short $r_s$ lengthscales? For the real galaxies 
considered in this paper, $r_{\rm s}$ are deduced from high-resolution optical 
and radio data, so their values are quite accurate. For distant galaxies, 
where such measurements are more difficult, the error must be evaluated 
theoretically and, conservatively, it may be a factor of $\mincir$2.
\smallskip

\noindent 
$\bullet$ Published estimates of the {\it energy of a core-collapse SN available 
for Galactic CR acceleration}, $(0.5-1)\,10^{50}$ erg, are fairly robust -- within 
a factor $\mincir$$2$ (Lingenfelter et al. 1998; Higdon et al. 1998; Tatischeff 
2008). This is $\sim$$5\%-10\%$ of the total kinetic energy of the SN ejecta, 
$\sim$$10^{51}$ erg (Woosley \& Weaver 1995), that is also a commonly assumed 
nominal value for this parameter in all core-collapse SNe. We assume this 
'Milky Way normalization' to hold also for the starburst galaxies considered in this paper. 
The observational fact that the much more massive 
($\magcir$25\,$M_\odot$) Wolf-Rayet stars seem to explode with 
$E_{\rm ej}$$\sim$$10^{52}$\,ergs (e.g., Lozinskaya \& Moiseev 2007) is a 
subtlety that, although important, does not substantially alter the general 
picture given the relative rarity of these stars.

We conclude that the quantities of Table\,1 galaxies relevant to 
Eq.(\ref{eq:CRp_density}) are quoted in the literature as observationally 
precise, to within a factor of $\mincir$2. In this case, the discrepancy 
between the very high CR energy densities deduced for active starburst nuclei 
($U_{\rm p}$$\approx$$100$\,eV\,cm$^{-3}$) and the low value deduced for 
the nuclear region of the Milky Way ($U_{\rm p}$$\approx$$5$\,eV\,cm$^{-3}$) 
seems very significant. Indeed, within our sample this discrepancy is 
especially striking between NGC\,253 and the Milky Way, which share very 
similar TIR luminosities and thus star formation rates.

\section{Supernova rates in star-forming environments}

In this section we investigate the constraints on the stellar initial mass 
function (IMF) that could possibly be derived from the SN rates that have 
been observed or deduced for our sample galaxies. 

A galactic SN rate can be modelled once we know {\it i)} the total gas mass 
that is turned into stars each year (i.e., the SFR), and {\it ii)} the fraction 
of young stars that have $M > 8\,M_\odot$ and are therefore SN progenitors. 

To evaluate the SFR we use Kennicutt's (1998a) popular $L_{\rm TIR} 
\rightarrow {\rm SFR}$ conversion:  
\begin{eqnarray}
{\rm SFR} ~=~ {L_{\rm TIR} \over 2.2 \times 10^{43} ~ {\rm erg} } ~~~~~ M_\odot\,{\rm yr}^{-1}\,,
\label{eq:kennicutt}
\end{eqnarray}
where $L_{\rm TIR}$ the total IR [i.e., $(8-1000)\mu$m] luminosity. In turn, the TIR flux is given by 
\begin{eqnarray}
\lefteqn{
f_{\rm TIR} \,=\, 1.8 \times 10^{-11} ~ [ 13.48\, f_{12} + 5.16\, f_{25} + 2.58\, f_{60} + f_{100} ]}
                \nonumber\\
& & {} ~~~~~~~~~~~~~~~~~~~~~~~~~~~~~~~~~~ {\rm erg~cm^{-2}~s}^{-1}
\label{eq:TIR}
\end{eqnarray}
(e.g., Sanders \& Mirabel 1996), where $f_\lambda$ are the {\it IRAS} spectral 
densities (in Jy) at $\lambda$ (in $\mu$m). The continuous-burst model of 
(10-100)\,Myr duration of Leitherer \& Heckman (1995) and the Salpeter stellar 
IMF with mass limits (0.1-100)\,$M_\odot$, provides a $L_{\rm TIR} \rightarrow 
{\rm SFR}$ conversion that applies most exactly to young starbursts embedded in optically 
thick dust clouds, such as the starburst regions in our sample galaxies. The calibration 
of Kennicutt's conversion has a $\sim$30$\%$ uncertainty that lies within the 
range of published calibrations (see Kennicutt 1998b)
\footnote{
	Observationally, the instantaneous IMF in a starburst is possibly different 
	from the time-integrated IMF (see review by Elmegreen 2005). Basically, 
	because of the proto-stars formed in an instantaneous burst of star 
	formation, the low-mass ones take a very long time to get to the main 
	sequence, and by the time they do, the upper-main-sequence stars have 
	already left the main sequence. This problem, of varying importance 
	according to the type of environment being considered, can affect the 
	calibration of Eq.(\ref{eq:kennicutt}) (see discussion in Kennicutt 1998b).
 }. 
In Table\,1 we report all relevant values for NGC\,253, M\,82, and Arp\,220; the 
corresponding SFRs are, respectively, 4.5, 10.2, and $220\,M_\odot$\,yr$^{-1}$. 
By comparison, our Galaxy has $L_{\rm TIR} \sim {1 \over 2}\,L_{\rm bol} 
\approx 5 \times 10^{43}$\,erg\,s$^{-1}$ and SFR$\,\approx\,2\,M_\odot$\,yr$^{-1}$. 

If the SN progenitors are all stars with $M$$\geq$8$\,M_\odot$, the gas mass that 
is turned into SN progenitors is 
\begin{eqnarray}
\dot M_{\rm gas}(\geq 8 M_\odot) ~=~ {\rm SFR} \times 
{\int_{8M_\odot}^{M_{\rm up}} m^{-x+1} ~ {\rm d}m 
\over  \int_{M_{\rm low}}^{M_{\rm up}} m^{-x+1} ~{\rm d}m} 
\end{eqnarray}
where $M_{\rm low}$ and $M_{\rm up}$ are, respectively, the lower and upper mass 
cutoffs of the stellar mass function, assumed here to be power-law with index $x$. 

Mass conservation requires that the gas mass turned into SN progenitors per year 
must equal the mass contained in $>$8$\,M_\odot$ stars formed per year, 
\begin{eqnarray}
\dot M_\star(\geq 8 M_\odot) ~=~ A ~\int_{8 M_\odot}^{M_{\rm up}} m^{-x+1}\, {\rm d}m
\end{eqnarray}
where $A$ is the normalization of the mass spectrum of the stars formed per year. 
By equating the two mass rates we can determine the value of $A$. We can then 
compute the number of SN progenitors that are formed per year, which in steady 
state is equal to the number of SN explosions per year, 
${\rm d}N_{\rm SN}/{\rm d}t = A \int_{8M_\odot}^{M_{\rm up}} m^{-x} {\rm d}m$, 
that is 
\begin{eqnarray}
\nu_{\rm SN} ~\equiv~ {{\rm d}N_{\rm SN} \over {\rm d}t} ~=~ 
{L_{\rm TIR} \over 2.2 \times 10^{43}} ~
{ \int_{8 M_\odot}^{M_{\rm up}} m^{-x}\, {\rm d}m \over 
\int_{M_{\rm low}}^{M_{\rm up}} m^{-x+1}\, {\rm d}m }  \,.
\label{eq:SNrate2}
\end{eqnarray}

In the evaluation of Eq.(\ref{eq:SNrate2}) we assume a Salpeter (1955) stellar IMF, 
i.e. $x=2.35$. The standard stellar mass range is usually taken to be bracketed 
by $M_{\rm low} \sim 0.1 \,M_\odot$ and $M_{\rm up}\sim 100\,M_\odot$ (e.g., 
Tinsley 1980). The two mass limits are dictated by considerations of ignition 
and sustainability of nuclear reactions ($M_{\rm low}$) and of stability to 
radiation pressure ($M_{\rm up}$). In what follows the exact value of the 
latter is relatively unimportant, whereas the value of the former is of 
great importance and hence will be discussed in some detail with reference 
to the physical status of the medium in the local star-forming environment.

In a non-turbulent environment, arguments of gravitational instability, efficient 
heat dissipation on collapse timescale, and thermal radiation of the collapsing 
object lead to the fragmentation limit, 
\begin{eqnarray}
M_{\rm frag} = 6 \mu^{-2} m_p 
\biggl({k_BT \over \mu m_pc^2}\biggr)^{1/4} 
\biggl({\hbar c \over Gm_p^2}\biggr)^{1/4}
\label{eq:fragm}
\end{eqnarray}
(e.g., Peacock 1999). For $T \sim 100\,K$, it is $M_{\rm frag} = 0.01\,
M_\odot$, almost an order of magnitude lower than the hydrogen-burning 
limit. (Brown dwarfs and planets may form with $M_{\rm frag} \leq M \leq 
M_{\rm burn}$, of course.) 

In a starburst environment, the (molecular gas) temperature is higher and the turbulence 
stronger. Because of the $T^{1/4}$ behavior in Eq.(\ref{eq:fragm}), the higher 
temperature causes only a modest increase of $M_{\rm frag}$. However, the substantial 
turbulence (comparable in energy to the gravitational energy) in the 
highly inhomogenous environment, where all the stars form, explode, drive shocks, 
accelerate cosmic rays and drive a dynamo (McKee \& Ostriker 1977), causes dense 
cores to be formed by turbulent shocks, and core sizes to scale with the thickness 
of post-shock gas, rather than with the local Jeans length. Therefore, fragmenting 
cores, much larger than the local fragmentation limit, can be assembled dynamically; 
indeed, the mass spectrum of dense proto-stellar cores generated by turbulent 
fragmentation is Salpeter-like for $M \magcir (1-2)\,M_\odot$ (Padoan \& Nordlund 
2002). The case for a top-heavy stellar IMF in a warm, dense, turbulent medium, 
due to a larger Jean mass, was also made by Klessen et al. (2007).

Therefore in evaluating Eq.(\ref{eq:SNrate2}) we assume $M_{\rm up}=100\,M_\odot$ 
and $M_{\rm low}=2\,M_\odot$. These assumptions yield 
\begin{eqnarray}
\nu_{\rm SN} ~ = ~
\left\{
\begin{array}{ll} ~ 0.12\,{\rm yr}^{-1} & \mbox{$~~~$ NGC\,253} \\
		  ~ 0.26\,{\rm yr}^{-1} & \mbox{$~~~$ M\,82} \\
		  ~ 5.6\,{\rm yr}^{-1} & \mbox{$~~~$ Arp\,220 } 
\end{array} 
\right.
\label{eq:SNrate3}
\end{eqnarray}
(For $M_{\rm low}=3\,M_\odot$, the estimated $\nu_{\rm SN}$ values are, 
respectively, 0.14, 0.32, 6.8\,yr$^{-1}$.) 
These estimates are in excellent agreement with observational values: 
$\nu_{\rm SN} \simeq (0.1-0.2)$\,yr$^{-1}$ (Antonucci \& Ulvestad 1988; Rieke 
et al. 1980, 1988), and certainly $\nu_{\rm SN} <0.3$\,yr$^{-1}$ (Ulvestad 
\& Antonucci 1997), for NGC\,253; 
$\nu_{\rm SN} \simeq (0.2-0.3)$\,yr$^{-1}$ (Kronberg et al. 1985; Rieke et al. 
1980) for M\,82; and 
$\nu_{\rm SN} \simeq (4 \pm 2)$\,yr$^{-1}$ (Lonsdale et al. 2006) for Arp\,220.
Using the 'standard' low value $M_{\rm low}=0.1\,M_\odot$, we would instead get 
$\nu_{\rm SN}$$=$ 0.03, 0.08, 1.6 \,yr$^{-1}$, which seems to be inconsistent 
with observations.

The above result 
suggests that in the central starbursts of M\,82, NGC\,253 and Arp\,220, the observed 
SN rates are consistent with the respective TIR luminosities (which are proxies 
for their SFRs), if their stellar IMFs (assumed to be Salpeter-like) are 
"top-heavy" with the lower mass cutoff of $M_{\rm low} \sim (2-3)\,M_\odot$. 
This deduction rings in with earlier suggestions on a high $M_{\rm low}$ 
truncation (at $\sim$$3\,M_\odot$) of the IMF in M\,82's nuclear starburst, based 
on the colors and luminosity of the galaxy (Rieke et al. 1980, 1993; Doane \& 
Mathews 1993). Top-heavy IMFs have been observationally suggested for 
Arp\,220 (Parra et al. 2007) and IC\,694 (P\'erez-Torres et al. 2009) based on 
the unusually large fraction of core-collapse SNe observed in these galaxies.
In comparison, in the Galaxy the 'standard' low-mass cutoff $M_{\rm low}= 
0.1\,M_\odot$ applies. With this low mass cutoff and $L_{\rm TIR} \sim 5 \times 
10^{43}$\,erg\,s$^{-1}$ in Eq.(\ref{eq:kennicutt}), the resulting predicted SN rate 
is $\nu_{\rm SN} \sim 0.015$\,yr$^{-1}$, in agreement with the observational 
value of $\nu_{\rm SN} \simeq (0.02 \pm 0.01)$\,yr$^{-1}$ (Diehl et al. 2006). 
In the LMC, too, the 'standard' low-mass cutoff seems to apply; using 
$M_{\rm low}=0.1\,M_\odot$ and SFR$\approx$0.25\,$M_\odot$yr$^{-1}$ (from TIR, 
thermal radio, and H$\alpha$ luminosities; e.g., Hughes et al. 2007 and Whitney 
et al. 2008), the predicted SN rate in the LMC is $\nu_{\rm SN}$$\approx$0.002\,yr$^{-1}$, 
in agreement with the observationally deduced value (van den Bergh \& Tammann 1991). 

An immediate consequence of a top-heavy stellar IMF is that the SFR as deduced 
from the TIR luminosity according to Eq.(\ref{eq:kennicutt}) is overestimated 
with respect to its real value. The degree of overestimation depends on the 
detailed shape of the actual stellar IMF. As an illustration, consider a stellar 
IMF defined as $\phi(m)$$=$$0$ for $0.1\,M_\odot$$\leq$$m$$<$$M_{\rm low}$ 
and $\phi(m)$$\propto$$m^{-2.35}$ (i.e., Salpeter-like) for 
$M_{\rm low}$$\leq m$$\leq 100\,M_\odot$, then the {\it real} SFR would 
be only a fraction $\int_{M_{\rm low}}^{100\,M_\odot} m^{-x+1}\, {\rm d}m / 
\int_{0.1\,M_\odot}^{100\,M_\odot} m^{-x+1}\, {\rm d}m $ of the {\it nominal} 
(i.e., Kennicutt-based) SFR: e.g., $M_{\rm low}$$=$$2\,M_\odot$ would entail 
SFR(real)$\,=\,$0.29\,$\times$\,SFR(nominal). On the other hand, if more 
realistically (e.g., Padoan \& Nordlund 2002) the IMF has a Salpeter-like 
profile for $M_{\rm low}$$\geq$$m$$\geq$$100\,M_\odot$ but it turns over at 
$m$$\sim$$M_{\rm low}$ such that $\phi(m)$$\propto$$m^{\alpha}$ (with 
$\alpha$$\geq$$0$) for $0.1\,M_\odot$$\leq$$m$$<$$M_{\rm low}$, then the real SFR 
would only be a fraction $f$$\equiv$$[\int_{M_{\rm low}}^{100\,M_\odot} m^{1-x} 
{\rm d}m / \int_{0.1\,M_\odot}^{100\,M_\odot} m^{1-x} {\rm d}m ] + [ M_{\rm 
low}^{-x} / \int_{0.1\,M_\odot}^{100\,M_\odot} m^{1-x} {\rm d}m ] ~ 
\int_{0.1\,M_\odot}^{M_{\rm low}} m^{\alpha+1} {\rm d}m$ of the nominal 
SFR; with $M_{\rm low}$$=$$2\,M_\odot$ it is $f$$=$$0.35, 0.36, 0.37, 0.42$ 
for $\alpha$$=$$0, 0.5, 1, 2$.

The realization that the SFR deduced from Eq.(\ref{eq:kennicutt}) may be overstimated 
does not, however, invalidate our estimation of the SN rates in starburst environments. 
Underlying Eq.(\ref{eq:kennicutt}) is the assumption of a Salpeter IMF for the stars 
in the mass range relevant to heating dust clouds (i.e., stars with $m \sim 
5\,M_\odot$ which emit copious UV radiation, where the absorption cross-section 
of dust grains peaks), which provides the correct normalization of the massive star 
formation rate. Only as a further step 
an extrapolation down to $0.1\,M_\odot$ is made 
in order to obtain the nominal SFR($\geq$$0.1M_\odot$) 
given by Eq.(\ref{eq:kennicutt}). Using a different calibration, say 
SFR($\geq$$5\,M_\odot$) as is common practice with 
radio SFR indicators (e.g., Condon 1992), this numerical issue would not show up.
However, a revision -- by a factor of $\sim$$3-4$ down -- of the SFR in the actively 
star forming galaxies at redshifts $z \magcir 1$, when the cosmic SFR was much higher 
than the current average, would have a major impact on our quantitative understanding 
of the star formation history of the Universe (e.g., van Dokkum 2008).

In summary, assuming the stellar IMF to be universal (Salpeter-like in the 
present calculation), we can model the fraction of newly formed SN progenitors 
($M \geq 8\,M_\odot$) by varying the low-mass cutoff from the 'canonical' value 
of $0.1\,M_\odot$, appropriate for a quiet environment like our Galaxy, to the 
'top-heavy' value of $(2-3)\,M_\odot$, suggested for a turbulent starburst environment. 
For a given gas mass turned into stars per year in a galaxy, the fraction of SN 
progenitors can then vary by a factor of $\sim$$4$. This variation is quite 
substantial, apparently well above the uncertainty in the quantities that enter 
Eq.(\ref{eq:CRp_density}).

\section{Conclusion}

The idea of a link between star formation and CR particles has been around 
for a long time. It is based on the recognition that the CR {\it electrons}, 
produced in the sites of SNe explosions (marking the end of $>$8$M_\odot$ 
stars progenitor stars), were responsible for diffuse synchrotron emission. 
The rough agreement between the relatively short lifetime ($\mincir$3$ 
\times$10$^7$yr) of these massive stars, and the similarly short synchrotron 
energy loss time of the electrons, led to the expectation that the non-thermal 
radio emission of a galaxy is a measure of its star formation activity on 
scales much shorter than the Hubble time. Biermann (1976) first included the 
non-thermal radio emission, $L_{\rm R}$, in quantitative galaxy population 
models. A $L_{\rm R}$--$\nu_{\rm SN}$ relation was given, theoretically, by 
Ulvestad (1982) and later, empirically, by Condon \& Yin (1990). A starburst model, 
that unified features of the earlier models and used $L_{\rm FIR}$ as the best 
estimate of the bolometric luminosity of a starburst, was proposed by Gehrz et al. 
(1983). Finally Condon (1992), relating $L_{\rm R}$ and $L_{\rm FIR}$ by 
attributing both luminosities to massive ($\geq$5$\,M_\odot$) stars, 
proposed a widely used, although semiempirical (e.g., no mechanism for CR 
generation in SN remnants, nor CR propagation, are explained), radio SFR 
indicator (see Persic \& Rephaeli 2007 and references therein for further 
developments).

The tight FIR/radio correlation (e.g., Condon 1992 and references therein) 
has been interpreted as a manifestation of the (apparently universal) link 
between non-thermal radio emission and recent star formation in galaxies. 
In particular, if the dust contributing to $L_{\rm FIR}$ is primarily heated 
by $\geq$5$M_\odot$ stars (Devereux \& Young 1990), if the CR energy production 
per SN is the same in all galaxies -- starburst and normal alike -- (V\"olk et al. 
1989), and if the magnetic field strength either plays a minor role or it 
varies little between galaxies (whether ordinary disk galaxies or luminous 
nuclear starbursts), then both $L_{\rm FIR}$ and the CR production rate are directly 
proportional to the recent massive SFR. 

Finally, the use of $L_{\rm FIR}$ (which in this paper is denoted by $L_{\rm 
IR}$) as an indicator of the current SFR has been long debated. The conversion 
is derived using synthetic population models. In the optically thick limit, it 
is only necessary to model the bolometric luminosity of the stellar population. 
In general, the choice of a suitable timescale for the stellar population (age 
of the starburst, timescale for sputtering of dust) will affect the calibration of the 
conversion: different assumptions by different authors have led to slightly 
different calibrations (see discussion by Kennicutt 1998b). For example, the 
conversion in Eq.(\ref{eq:kennicutt}) applies only to young, optically thick starbursts; 
whereas in more quiescent normal star-forming galaxies the relation is more 
complicated, due to the increasing contribution to dust heating from from old 
stars, and the decreasing dust optical depth. 

In the present paper we have explored the link between star formation and CR 
{\it protons}; specifically, the link between ongoing star formation and $U_{\rm 
p}$. Doing this we have followed up, and to some extent ultimately strengthened, 
the longtime suggestion that CRs are related to SN explosions (Ginzburg \& 
Syrovatskii 1964; see also Salvati \& Sacco 2008). 

We started by assuming that equipartition arguments provide realistic estimates of 
the particle energy densities, $U_{\rm p}$, in some well-known starbursts with measured 
synchrotron radio emission. We then found that in these environments, for a given gas 
mass per year turned into stars (measured by $L_{\rm TIR}$), a "top-heavy" stellar 
IMF with a low-mass cutoff of $\sim$$2\,M_\odot$ (as befits protostar fragmentation 
in a turbulent medium) leads to the nominal SN rate required to sustain the correct 
$U_{\rm p}$. By contrast in our Galaxy and in the LMC, where star formation is much more moderate, 
the (measured) $U_{\rm p}$ is consistent with the standard ($M_{\rm low} = 
0.1\,M_\odot$) IMF. In general, we may conclude that the instantaneous IMF 
in a starburst environment is possibly different from the time-integrated IMF (the latter 
being represented by the standard Salpeter function).

A top-heavy stellar IMF should have appreciable effects on galaxy colours. For a 
given $\dot M_{\rm gas}$, hot massive OB stars are more numerous, so the IR hump 
in the galaxy's broad-band continuum is enhanced, if the IMF is top-heavy rather 
than standard. Starburst galaxies are then expected to turn up more IR-bright than they 
would be with a standard IMF. Indeed, a detailed analysis of the colors of M\,82 
and NGC\,253 led Rieke et al. (1980, 1993) and Doane \& Mathews (1993) to suggest 
that the stellar IMF there was truncated at $M_{\rm low} \simeq 3\,M_\odot$. 

In each starburst galaxy considered in this paper, for the relevant quantities appearing 
in Eq.(\ref{eq:CRp_density}), we have used nominal values from the literature. 
These, although statistically consistent within a factor of $\mincir$$2$, may 
however be affected by systematics that are presently difficult to model, so the 
quantities appearing in Eq.(\ref{eq:CRp_density}) could be known more poorly than 
we may think. Given the only initial direct $\gamma$-ray--based determinations of 
$U_{\rm p}$, and the errors (statistical and systematic) affecting the relevant 
quantities ($\nu_{\rm SN}$, $\tau_-$, $\eta E_{\rm ej}$, $r_{\rm s}$), our present 
deductions on the stellar IMF are clearly only tentative. 

Indeed the question whether 
a low mass limit on the stellar IMF in a starburst region can be derived has been argued 
to be ultimately difficult (e.g., Elmegreen 2005), because we are probably missing 
some crucial subtlety in the time-integrated star formation. Perhaps a relatively 
robust result is that, due to the extra energy input available from the starburst environment, 
the timescale required for proto-stellar clouds to collapse, get rid of their angular 
momentum, split up, and merge again, is just very much longer within a starburst than outside 
the region. So the corresponding low-mass stars would presumably stay in the Hayashi 
phase for a time longer than the starburst, and would arrive on the main sequence only after 
the starburst activity has subsided.

In any case, as concerns the derivation of the massive SFR from CR protons, direct 
$\gamma$-ray detections of starburst galaxies (in the 0.1-100\,GeV photon energy range with 
the Large Area Telescope onboard the {\it Fermi} Gamma-ray Space Telescope; and in 
the $>$$100$\,GeV range with Cherenkov telescope arrays such as H.E.S.S., VERITAS, 
and MAGIC) have just started providing direct $U_{\rm p}$ measurements, which 
happen to be consistent with theoretical predictions, for at least some of 
the nearby galaxies. Given these initial agreements, should our predictions on 
$U_{\rm p}$ be more generally confirmed, some immediate implications would be: 
\smallskip

\noindent
{\it (i)} star forming galaxies are powerful particle accelerators, able to 
achieve CR energy densities orders of magnitude higher than what is measured 
in the Milky Way and in the LMC; 
\smallskip

\noindent
{\it (ii)} SNe, both in quietly star-forming galaxies like the 
Milky Way and the LMC and in very actively star-forming galaxies such as those 
considered in this paper, probably share a common (universal?) CR acceleration 
efficiency;    
\smallskip

\noindent
{\it (iii)} particle energy densities and equipartition magnetic fields can be 
used (at least statistically) as proxies of the actual quantities; 
this could be particularly useful in the case of galaxies that are too far for 
their (unbeamed) $\gamma$-ray emission to be measured; 
\smallskip

\noindent
{\it (iv)} 
the initial stellar mass spectrum shows a systematic dependence on the local 
star-forming environment (see also, e.g., Elmegreen 2005); and  
\smallskip

\noindent
{\it (v)} contrary to common practice, each given SFR indicator should have 
different calibrations according to whether galaxy luminosities are dominated 
by the {\it current} or the {\it average past} SFR -- the former case including 
local starburst galaxies, ultra-luminous IR galaxies (ULIRGs), and high-redshift galaxies; 
the latter case including most local galaxies (e.g., Persic \& Rephaeli 2007). 
If so, since in the earlier Universe star formation took place in denser 
environments than at the present epoch, the calibration of SFR indicators would 
show a systematic dependence on redshift. This would deeply affect our 
understanding of the star formation history of the Universe (van Dokkum 2008).
\medskip

\noindent
{\it Acknowledgements.} Useful remarks by an anonymous referee are gratefully 
acknowledged. We dedicate this paper to the memory of our colleague and friend 
Elihu A. Boldt.

\end{document}